# PRESSURE FLUCTUATIONS, VISCOSITY, AND BROWNIAN MOTION

Frank Munley
Roanoke College, Salem, VA 24153

## ABSTRACT

Brownian motion occurs in a variety of fluids, from rare gases to liquids. The Langevin equation, describing friction and agitation forces in statistical balance, is one of the most successful ways to treat the phenomenon.

In rare gases, it is appropriate to model both friction and agitation in terms of independent molecular impacts with the particle. But in relatively dense fluids, such as water and air at standard temperature and pressure, the mean free path between collisions of fluid molecules is much smaller than the size of the Brownian particle, and the friction is normally treated as a mesoscopic viscous effect described by Stokes' Law which treats the fluid as continuous. The appropriateness of using Stokes' Law will be discussed in terms of recent experimental research in the ballistic or "coasting" phase of motion occurring at a very short time scale.

Given the mesoscopic nature of the friction force for relatively dense fluids, we should expect the agitation force to also be mesoscopic. But it is often unrealistically modeled as uncorrelated individual impacts. It has been suggested occasionally that mesoscopic pressure fluctuations are appropriate for denser fluids. The purpose of this paper is to model friction as a result of mesoscopic pressure fluctuations. First, the simple random walk will be used to approximate the time and space scales below which ballistic motion begins and diffusive motion ends. Following that, pressure fluctuations and the associated time scale will be introduced to explain Brownian motion.

As successful as the pressure fluctuation model is for many fluids, it will be shown to fail for fluids like glycerin that have viscosities a thousand times and more that of water.

## I. INTRODUCTION

Brownian motion is the incessant irregular motion of small particles suspended in a fluid. Robert Brown considered the phenomenon in depth in 1827, following earlier observations by several other investigators.[1] Brownian motion is observed with pollen grains as in Brown's original experiments, or with milk fat cells, smoke particles, or other micron-size particles. The modern choice is engineered polystyrene beads of uniform size. For convenience, the particles must be small enough for motion to be easily observed in an optical microscope, typically $10^{-6}$ m or smaller. More advanced techniques, such as Mössbauer spectroscopy[2] and photon correlation spectroscopy,[3] are able to observe Brownian properties of even smaller particles.

Albert Einstein provided a firm theoretical basis for the phenomenon in 1905.[4] His approach exploited an analogy between diffusion of solute molecules in a fluid solvent and the diffusion of the much larger Brownian particles (BP for short) in a fluid. The analogy today seems compelling and even obvious, but in Einstein's day it was a bold one for two reasons. First, diffusion was seen only as a phenomenon in which molecules of more or less equal size and mass interact, so each collision between two molecules has a large effect on both. How could a series of single impacts of very tiny molecules cause a much larger BP to move significant distances? Wouldn't the small impacts occurring in random directions cancel out?[5] And second, some eminent chemists and physicists at the time, such as Ostwald and Mach, still doubted to various degrees the existence of atoms and molecules, considering them to be just a convenient construction to help explain chemical reactions and physical properties of materials and therefore not a real explanation of Brownian motion.

Einstein also provided an alternative "random walk" model which assumes the particle takes steps of finite length in random directions, the changes in direction resulting from a random agitation force acting on the BP.[6] The diffusion equation then results in the limit of rapid, very short steps. The random jaggedness of the particle path, though only an artifact of connecting successive plot points of periodic observations of a Brownian particle, correctly leads to the conclusion that the average squared displacement of a Brownian particle from the starting point is proportional to time, and the dispersion of a large number of particles all starting from the same point is bell-shaped, i.e., Gaussian, the width of the Gaussian increasing linearly with time. This dependence, plus other characteristics of the motion based on the atomic theory of matter and statistical methods Einstein himself devised, provided the basis for the experimental verification of his theory by Jean Perrin in a number of experimental investigations published in 1908-1909. Perrin's work convinced most if not all skeptics that the atomic composition of matter was indeed a reality and not just a contrivance to understand chemical reactions.

In 1908, Paul Langevin proposed a theory of Brownian motion based on Newton's second law of motion in which two fluid forces guide the particle's motion.[7] First, as the BP drifts through the fluid, it is subject to a frictional or drag force on it that alone would bring it to a stop and erase any memory of its original motion. And second, there is fluctuating thermal agitation from the fluid that sustains the motion by providing impulses in random directions to the BP. The two processes together—the random impulses keeping the particle in motion and the systematic friction erasing the motion—tend to balance each other so the instantaneous speed of the BP is, on average, the thermal speed one would expect from the equipartition theorem of statistical mechanics requiring the average kinetic energy to be proportional to temperature.[8] (See Eq. (7) below.) In this way, thermal equilibrium is maintained.

Investigations of Brownian motion did not stop with Langevin and common liquids like air and water. Today, research centers on novel materials consisting of traps that temporarily retard Brownian particles; on soft materials; and on hydrodynamic subtleties affecting motion which lead to anomalous Brownian motion that exhibit non-Gaussian dispersions.[9]

There are important similarities and differences between various fluids in their friction and agitation mechanisms. For example, the friction forces on a BP in both very low-density (rarified) gases and denser fluids (like air and water) are proportional to particle speed. (This is quite unlike the friction force on, say, a pitched baseball where it is proportional to the square of the speed.) But these forces differ in their dependence on BP size, in the former case being proportional to the cross-sectional area of the BP and in the latter to the linear dimension (e.g., the radius of a spherical BP). Another important difference is the molecular nature of the friction: in low-density gases, it results from random independent impacts of individual molecules, while in dense fluids the friction is viscous in nature, involving the *mesoscopic collective* action of colliding fluid molecules surrounding the BP. In other words, for Stokes friction in a dense fluid, molecular impacts on the BP surface are the cause of the agitation force only in the trivial sense that they convey to the particle collective physical effects originating on a wider scale from the interactions between colliding molecules. For rarified gases, on the other hand, intermolecular collisions are irrelevant to the frictional effect.

In light of the collective nature of Stokes friction in a dense fluid, it is only natural to expect the agitation term to also be of a collective nature arising from mesoscopic processes on the spatial scale of the Brownian particle. F.L. Markley and D. Park made that point in 1972[10] and provided the motivation for the present work, remarking:

"[Brownian motion] must be considered to be a macroscopic rather than a microscopic phenomenon. We can think of it as caused by pressure fluctuations due to statistical fluctuations in the number of molecules in a volume of order of magnitude of the volume of the Brownian particle."

A month later, Kivelson *et al.* presented a theory of the role of pressure fluctuations in Brownian motion.[11] Their treatment, which focused more on the reliability of Stokes' Law than on the agitation process *per se*, relied on derivations of fluid properties using complex statistical mechanics. These considerations lead to the main purpose of the paper: to explain, in a simple accessible way, the role of pressure fluctuations as the proper *mesoscopic* counterpart to viscous friction in denser fluids. Pressure fluctuations are characterized by a well-defined time scale, thus allowing an alternative to the "white noise" or Wiener process which has an undefined time scale often taken to be infinitesimal. Grassia gives a very clear treatment of the white noise formulation, and adapts it for finite time intervals useful for fluctuations of finite duration in a low-density gas.[12]

In Section II, the random walk model will be used to illustrate the relationship between the experimentally measurable diffusion constant and the step size between periodically observed particle positions of the random walk process. This will lead to estimates of the minimum step and time scales possible for a random walk model, shorter times leading to smooth ballistic motion. Water and air, and a Brownian particle of radius $10^{-6}$ m with twice the density of water, will be used to illustrate minimum values. In Section III, the Langevin equation will be introduced and the conditions needed for the oft-used Stokes (or Stokes-Einstein) friction will be discussed. Pressure fluctuations will be introduced in Section IV and an intuitive approximation of the fluctuation-dissipation theorem given. The theorem will be used to verify expressions for time scale and pressure fluctuations. Section V shows why the pressure fluctuation mechanism fails for extremely viscous substances such as glycerin, where somewhat paradoxically it takes a shorter time for a particle at rest to be boosted to its thermal speed. A modification of the Langevin equation will be proposed to deal with high-viscosity liquids like glycerin. Section VI will summarize the results.

## II. RANDOM WALK, BALLISTIC MOTION, MINIMUM TIME AND SPACE SCALES

A random walk picture of Brownian motion can be constructed when a BP under a microscope diffuses from its initial position. Visually, the BP undergoes a tiny jiggling motion so rapid it is impossible to follow with the eye in detail. But single-particle positions can be observed and recorded periodically at equal intervals of time, during which the particle noticeably changes position. When successive positions are plotted and connected by points, a jagged "random walk" is contrived as a visualization of the underlying physical process. For example, some of the famous experiments of Jean Perrin in 1908-1909 recorded a single particle's position in water every 30 seconds.[13] Perrin's particles were about $10^{-6}$ m in diameter, and on average a particle moved a few particle diameters between observations. During the time between observations, random agitation from the fluid results in the direction gradually changing. An idealized example of such a picture in Figure 1 shows the contrived jagged path of the particle. One of the steps is magnified to show what happens if another observer were to track the same particle at the same time, but recorded the particle's positions more frequently, say every few seconds. As shown, the jaggedness, but now with shorter steps, persists even over the shorter times.

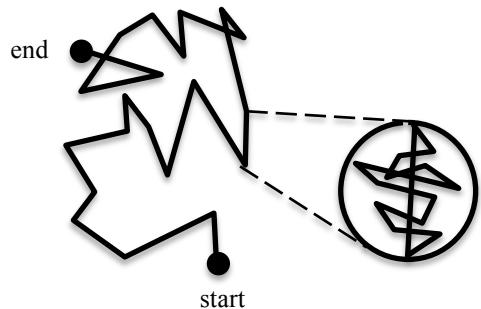

**Figure 1:** A random walk where the particle position is periodically observed, say every minute or so. The circle magnifies one step showing the extra detail if the particle is observed on a shorter time scale, say every 30 seconds or so.

Knowing the time between observations, say 30 s as in Perrin's experiments and imagined in Figure 1, the particle's average speed on segments between consecutively observed positions can be calculated. But as the magnified step in Figure 1 shows, if a smaller period between observations shows the jaggedness to persist along the smaller steps, the distance moved between two end points 30 s apart in time is really greater than would be calculated just using the straight-line distance between those two end points. Consequently, the particle's speed in traveling between two points is really greater than that calculated with a 30 s between observations. Suppose now a third observer had recorded positions with an even shorter period, say a few tenths of a second. Then if the jaggedness persists, but now on an even smaller scale, the distance the particle moved in 30 s is again bigger than originally thought, and its speed is again greater than thought. Clearly, since a particle can't move faster than the speed of light (in fact, it moves much slower than that), *the jaggedness must stop at some minimum observation time*, typically about a microsecond or less, the motion smoothing out to "ballistic motion" for shorter observation times.[14] This is shown in Figure 2.

Despite the necessity of jaggedness giving way to smoothness, careless statements are sometimes made implying that the jaggedness continues to arbitrarily small time and space scales.[15] The reason for the misleading claim is that the white noise or Wiener process with its minimum time scale of zero can be useful for applications where precise values of minimum time and space scales are not of interest or are not easily modeled.

Ballistic motion was first observed in 2005.[16] Since then, a number of excellent experiments have been done for gases and liquids, the simplest result for the latter being in acetone which has a very low viscosity compared to water and many other liquids.[17] The true ballistic path is not absolutely smooth because of the quasi-continuous randomly acting agitation force, as suggested in the magnified portion of Figure 2(b). Individual distortions are fantastically small as we will see at the conclusion of Section IV, but like a typical random walk process, agitation impulses combine to produce larger changes in direction over longer times.

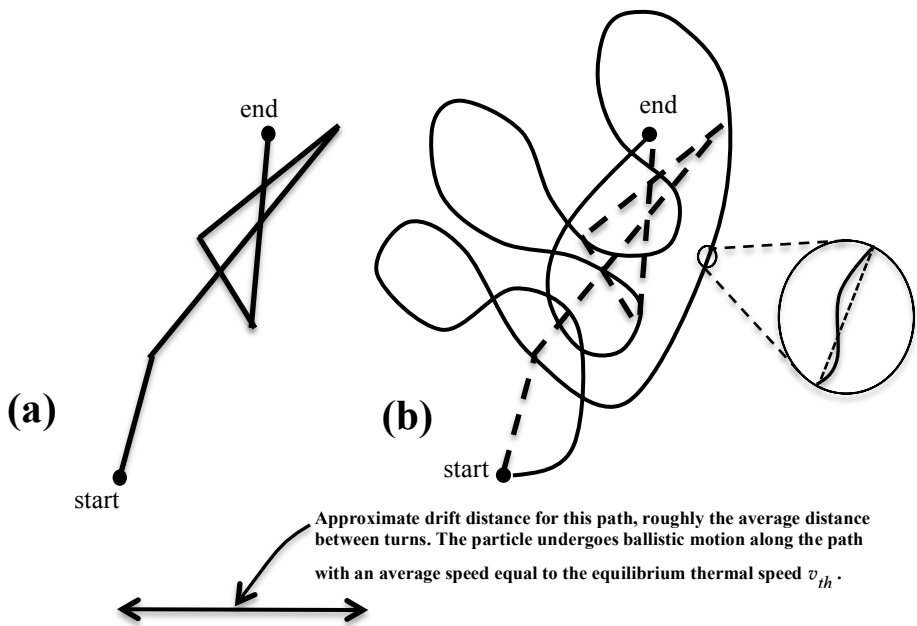

**Approximate drift distance for this path, roughly the average distance between turns. The particle undergoes ballistic motion along the path with an average speed equal to the equilibrium thermal speed $v_{th}$.**

**Figure 2:** **(a)** The solid line shows the random walk of a BP in a rare gas or denser fluids (like water or air at standard temperature and pressure (STP)), the straight lines being drawn between positions of the particles "strobed" at equal intervals of time. **(b)** The true continuous path of the same BP is almost perfectly smooth on a very small, essentially continuous, time scale. The magnified portion of a very tiny part of the curve shows over very short distances the slight, random drifts around the average direction of motion due to thermal impulses from the fluid that barely change the particle's direction of travel but combine over time and larger distances to gradually change the direction of motion. The smooth "ballistic" almost-straight-line drift motion indicated in **(b)** is observable only with a very short period between observations.

We will now see that good approximations to the minimum random walk scales are derivable from a simple model. The results follow from the very important property of a random walk, that the average *squared* distance a particle wanders from its starting point is proportional to the number of steps taken. This property is easily proved as follows.

Imagine a BP's position in three dimensions to be observed and recorded every $t_0$ seconds for a total time $t_n = nt_0$, with $t_0$ of the order of seconds so the number of steps is $n \gg 1$. The net vector displacement is:

$$\boldsymbol{r}(t_n) = \sum_{i=1}^{n} \boldsymbol{r}_i \ . \qquad (1)$$

The square of the displacement from the starting point after these $n$ steps taken in time $nt_0$ is

$$r^2(t_n) = \sum_{i=1}^{n} r_i^2 + \sum_{i \neq k}^{n} \boldsymbol{r}_i \cdot \boldsymbol{r}_k \approx \sum_{i=1}^{n} r_i^2 \ .$$ The approximation is valid for $n \gg 1$ because successive steps are random in direction. Now let the same procedure be carried out for a large number of particles so a good average value of the first sum can be obtained, giving $\left\langle r^2(t_n) \right\rangle = \left\langle \sum_{i=1}^{n} r_i^2 \right\rangle$.

Then the average squared single-step size, $\Delta_{t_0}^2$, is: $\Delta_{t_0}^2 = (1/n) \left\langle \sum_{i=1}^{n} r_i^2 \right\rangle = \left\langle r^2(t_n) \right\rangle / n$, so:

$$\left\langle r^2(t_n) \right\rangle = n\Delta_{t_0}^2, \text{ QED} \ . \qquad (2)$$

Assuming spatial isotropy, the squared average of displacement of projections along $x$, $y$, and $z$ are equal: $\Delta_{t_0 x}^2 = \Delta_{t_0 y}^2 = \Delta_{t_0 z}^2$, and $\Delta_{t_0}^2 = \Delta_{t_0 x}^2 + \Delta_{t_0 y}^2 + \Delta_{t_0 z}^2$. So if the data is recorded in d dimensions (d = 1, 2, or 3), then $\Delta_{t_0}^2 = d\Delta_{t_0 x}^2$, which leads to:

$$\left\langle r^2(t_n) \right\rangle = n d\Delta_{1x}^2 \ . \qquad (3)$$

For example, it is common to observe Brownian particles through a microscope, with positions recorded on a d = 2 grid system. Perrin published pictures of such two-dimensional trajectories, observations being made every 30 s as mentioned above. But his numerical analysis was based only on projections of the path segments on the $x$-axis, so d = 1.

In Eq. (3), replace $n$ by $t_n / t_0$. Then the diffusion parameter $D$ in the random walk context is defined from the following equation:

$$\left\langle r^2(t) \right\rangle = n\Delta_{t_0}^2 = n d\Delta_{t_0 x}^2 = (t / t_0) d\Delta_{t_0 x}^2 = 2dDt \ , \qquad (4)$$

the factor of 2 being an oft-used convention. Eq. (4) shows that the root mean square displacement $\left\langle r^2 \right\rangle$ goes as $t$, as we expect from random steps reflected in data obtained by observing every $t_0$ seconds. Solving for $D$, we get:

$$D = \frac{\Delta_{t_0 x}^2}{2t_0} = \frac{\Delta_{t_0}^2}{2dt_0} \ . \qquad (5)$$

Note that Eqs. (4) and (5) are *robust* results as long as the value of $t_0$ is not so small that ballistic motion is involved.

It might appear from Eq. (5) that $D$ depends on $t_0$ and $\Delta_{t_0}$ (or $\Delta_{t_{0x}}$), but in fact $D$ is a constant because $\Delta_{t_0}^2$ is proportional to $t_0$.[18] Another way to see the constancy is to note from Eq. (4) that the same particle can be simultaneously tracked for the same total time $t$ by two observers utilizing different values of $t_0$, and they must agree on the observed total displacement. Doing this for many particles, they get the same result for $\langle r^2 \rangle$, which determines $D$. The constancy of $D$ will appear again in the next section, where Eq. (17) expresses $D$ in terms of parameters which are physical constants or parameters of the experimental procedure such as particle mass and temperature.

Eq. (5) implies the speed problem mentioned above and suggested in Figure 1. To see this explicitly, note that the *apparent* average squared speed between two position observations along a straight-line segment of average squared length $\Delta_{t_0}^2$ is:

$$\langle v_{t_0}^2 \rangle = \left( \Delta_{t_0} / t_0 \right)^2 = \mathrm{d}\Delta_{t_0x}^2 / t_0^2 = 2D\,\mathrm{d} / t_0, \qquad (6)$$

where the final member of the equation substitutes for $\Delta_{t_0x}^2$ from Eq. (5). Therefore, since $D$ and d are constants, $\langle v_{t_0}^2 \rangle \rightarrow \infty$ as $t_0 \rightarrow 0$. So there must exist a minimum random walk time $t_{\min}$, below which the jagged path between observation times smooths out to the "ballistic motion" phase. And associated with $t_{\min}$ in the ballistic phase, there is an average three-dimensional distance $\Delta_{\min}$ over which the particle travels with a more-or-less constant velocity which changes only gradually as tiny random impulses affect its motion.

The random walk parameters $t_{\min}$ and $\Delta_{\min}$ below which the ballistic phase appears can be determined by imposing two conditions: (1) the speed of the particle over the ballistic path should on average be the thermal equilibrium speed required by the equipartition theorem:

$$\Delta_{\min} / t_{\min} = \sqrt{\mathrm{d}k_B T / M} = v_{th,\mathrm{d}}; \qquad (7)$$

and: (2) the expression for the diffusion constant in Eq. (5) must be obeyed for these values:

$$D = \frac{\Delta_{\min}^2}{2\mathrm{d}\,t_{\min}}. \qquad (8)$$

Solving Eqs. (7) and (8) for $t_{\min}$ and $\Delta_{\min}$, we get:

$$t_{\min} = \frac{2DM}{k_B T}, \qquad (9)$$

$$\Delta_{\min} = 2D\sqrt{\frac{\mathrm{d}M}{k_B T}}. \qquad (10)$$

To illustrate for water and d = 3, consider a BP with a radius of $a = 10^{-6}$ m and a density of twice that of water, so its mass is $M = 8.34\text{x}10^{-15}\text{kg}$. Then at room temperature (300 K), using the experimentally measured value $D = 2.2\text{x}10^{-13}\text{m}^2/\text{s}$, we find $\Delta_{\min} = 1.1\times10^{-9}\text{m}$ and $t_{min} = 8.8 \times 10^{-7}\,\text{s} \approx 10^{-6}\,\text{s}$ . So to observe ballistic motion, a strobe time $\ll 10^{-6}\,\text{s}$ is needed. For air, $D = 1.3 \times 10^{-11}\,\text{m}^2/\text{s}$ , and a similar calculation gives a strobe time $\ll 10^{-5}\,\text{s}$ .

## III. THE LANGEVIN EQUATION: GENERAL CONSIDERATIONS AND FRICTION FORCE

The random walk model helps to understand some of the basic parameters of Brownian motion, but the Langevin equation mentioned in Section I is the more precise, well-worn, and fruitful approach. For both low-density gases and for denser fluids like air and water, the frictional effect on a moving BP, under conditions to be discussed, is proportional to the BP's speed. Then assuming that forces external to the liquid are absent or irrelevant, the general form of the Langevin's version of Newton's second law for low-density gases and denser fluids is:

$$M \frac{d\boldsymbol{v}}{dt} = -f\boldsymbol{v} + \delta\boldsymbol{F}_a(t) \; . \qquad (11)$$

Here, $f$ is the friction constant, $\boldsymbol{v}$ is the BP velocity measured in the lab frame (where the fluid is at rest from a macroscopic perspective), $M$ is the BP mass, and $\delta\boldsymbol{F}_a(t)$ is the random force originating from the thermal action of the fluid. $\delta\boldsymbol{F}_a(t)$ is on average isotropic in the lab frame.

The linear addition of friction and agitation forces in Eq. (11) implies that they are independent, which is a key assumption of the Langevin approach. Physically, their independence is suggested by the fact that a BP initially not moving and so with zero friction force, will not stay that way but will start moving because of random thermal agitation forces from the fluid. The independence also turns on the assumption that the basic events causing the agitation occur on a much shorter time scale than the relaxation time, i.e., the time it takes friction to dissipate an initial motion. Consequently, the agitation force appears as a very gentle, almost constant, background force that slowly changes the direction of motion and sustains it as the friction force acts. The difference in time scale is obvious for a rarified gas, where a random series of individual molecular impacts agitate the particle, a single impact having a very small effect on the particle whose mass is $10^{12}$ times that of a molecular mass. Friction in this case also relies on individual molecular impacts *via* the "raindrop effect" whereby impacts are greater on the front side of a moving particle. Given the much slower thermal speed of the particle compared to the molecular speed, the asymmetry of impacts is small, making the relaxation time much larger than the time needed for a single impact to occur. Friction and agitation mechanisms are also independent for denser fluids like air and water, again because of the great difference in time scales for the two processes.

For a spherical BP of radius $a$, it is often most appropriate for denser fluids to use the well-known Stokes friction constant, $f_{Stokes}$ , which is proportional to $a$ rather than the cross-sectional area $\pi a^2$ appropriate for low-density gases:

$$f_{Stokes} \equiv f_{St} = 6\pi\eta a \; , \qquad (12)$$

where $\eta$ is the shear viscosity of the fluid. Stokes friction is an excellent representation of the frictional force under the following assumptions: (1) the mean free path of the fluid molecules is much smaller than the size of the particle; (2) the speed of the particle through the fluid is not too great; (3) the particle is not significantly accelerated, so the velocity of the particle through the fluid is approximately constant;[19] and (4) the "no slip" condition holds, i.e., the fluid molecules do not slide over the moving particle's surface but move along with it.

The mean free path for air at STP is about $6 \times 10^{-8}\,\text{m}$ and much shorter for water, so condition (1) is satisfied for particles about a micrometer in radius. In condition (2), "not too great" in the case of speed means that the Reynolds number is small. This dimensionless number, a ratio of inertial to viscous forces on a particle of linear dimension $a$ moving with speed $v$ through the fluid of density $\rho$, is:

$$Re = \frac{av\rho}{\eta}.\qquad(13)$$

When Re is $<<1$, viscous forces dominate and the fluid undergoes laminar (non-turbulent) flow around a particle moving at constant velocity. Since the average $v$ in Brownian motion is the thermal speed of the particle, it is a simple matter to show that Re $<<1$ for any conceivable combination of the parameters in Eq. (13) consistent with the mean free path of the molecules being much less than the particle size. Therefore, condition (2) is satisfied. Condition (3) is well satisfied in air, where accelerations are small and (in Nelson's words above) "unbelievably gentle." For liquids, this requirement is sufficiently satisfied in water and very well satisfied in a liquid of low viscosity (e.g., acetone) using particles of high density (e.g., barium titanate). Condition (4) is generally satisfied, and leads to excellent agreement with a variety of experiments with liquids. (The no-slip condition entrains fluid with the particle, giving the complex a mass greater than the particle mass, a fact accounted for in Eq. (12) when Condition (3) is satisfied.[20]) So $f_{St}$ in Eq. (12) an excellent approximation for a particle undergoing Brownian motion in fluids like air and water.[21]

Although the conditions (1)—(4) are sufficiently satisfied for the diffusion (long-time) phase of motion, Stokes friction is not precisely valid in the ballistic motion phase. In that phase, the particle is affected by a vortical type of motion set up in the fluid by the particle's motion through it. These vortices swirl around and then interact with the particle at a slightly later time, causing a memory effect in the motion and an increase in the particle's effective mass beyond that accompanying Stokes friction. This complex hydrodynamic action, first recognized and explained in the 1880s,[22] causes the ballistic motion to differ a bit from what it would be with an ideal Stokes friction. But for longer times, the simpler Stokes friction leads to the correct mean-square displacement and hence the correct diffusion constant, as shown in Eq. (4). This immunity of the long-time diffusion phase to the hydrodynamic effect's increase of the particle's effective mass is a consequence of $D$ being independent of the mass of the particle, which is explained in the discussion following Eq. (17) below. Since the ballistic phase is not the focus of this paper, the simpler Stokes friction will be used in what follows

Because the friction and agitation terms in the Langevin equation add linearly and act independently, the characteristic *relaxation time* for the friction term can be calculated by looking at the action of friction alone.[23] Assume a BP has the velocity $v_0$ at time $t = 0$. (For smooth motion as in Figure 1, the magnitude, $v_0$, of $v_0$ will be on average equal to the thermal speed $v_{th}$ in Eq. (7).) Neglecting the agitation term in Eq. (11), we obtain upon solving for $v(t)$:

$$v(t) = v_0 e^{-t/\tau},\qquad(14)$$

where the relaxation time $\tau$ due to friction is $\tau = M/f$, or for Stokes friction,

$$\tau = \frac{M}{6\pi\eta a}.\qquad(15)$$

As Eq. (14) shows, friction erases an original velocity gradually with relaxation time $\tau$, which is large compared to the much smaller agitation time. The small random changes in the particle velocity from agitation, each taken later as of constant magnitude $|\delta v_i| = \delta v$, add vectorially to $v(t)$ as suggested in Figure 3 below. Each random change also dissipates according to Eq. (14):

$$\delta v(t + t') = \delta v(t) e^{-t'/\tau}.$$

The erasure time $\tau$ in Eq. (15) should be comparable to the random walk characteristic time in Eq. (9). Substituting $\tau$ for $t_{rw}$ in Eq. (9) and solving for $D$, we get $D = \dfrac{k_B T}{12\pi \eta a}$. But as will now be shown, this is half the true value, showing the limitations of the simple random walk model when used for times comparable to or smaller than $t_{rw}$, where smooth ballistic motion, not diffusion, becomes important.

For a rigorous derivation of $D$,[24] dot both sides of Eq. (11) with $\boldsymbol{r}$, use the identities $\boldsymbol{r} \cdot \boldsymbol{v} = d^2 r^2 / 2 dt^2 - v^2$ and $d(r^2)/dt = 2\boldsymbol{r} \cdot (d\boldsymbol{r}/dt) = 2\boldsymbol{r} \cdot \boldsymbol{v}$. Taking averages and substituting into Eq. (11), we get:

$$\frac{M}{2} \frac{d^2 \langle r^2 \rangle}{dt^2} - M \langle v^2 \rangle = -\frac{f}{2} \frac{d \langle r^2 \rangle}{dt} + \langle \boldsymbol{r} \cdot \delta \boldsymbol{F}_a \rangle . \qquad (16)$$

But $\langle \boldsymbol{r} \cdot \delta \boldsymbol{F}_a \rangle = 0$,[25] leading to the solution: $\langle r^2 \rangle = C_1 e^{-t/\tau} + C_2 + C_3 t$. Assuming the particle starts at $\boldsymbol{r} = 0$, then $C_2 = 0$. And in the diffusion realm where $t \gg \tau$, the first term, $C_1 e^{-t/\tau}$, can be neglected. Then $\langle r^2 \rangle = C_3 t$. Substituting this into Eq. (16), we get $(f/2)C_3 = M \langle v^2 \rangle$, which gives $\langle r^2 \rangle = (2M \langle v^2 \rangle / f) t$. Substituting $d k_B T / M$ for $\langle v^2 \rangle$, we get the diffusion equation: $\langle r^2 \rangle = 2 d k_B T t / f \equiv 2 d D t$, showing that:

$$D = \frac{k_B T}{f} . \qquad (17)$$

For Stokes friction, $f_{St} = 6\pi \eta a$ as in Eq. (12), and $D$ depends on fluid viscosity, particle radius, but not on particle mass. This is true in general when $f \propto v$. This independence of $D$ on particle mass occurs because a bigger particle mass causes the relaxation time in Eq. (15) to increase, but a larger mass also has a smaller thermal speed as shown in Eq. (7), tending to decrease $D$, and the two tendencies cancel each other.

## IV. THE LANGEVIN EQUATION: PRESSURE FLUCTUATION AS AGITATION MECHANISM

Since we will focus on pressure fluctuations, we change the subscripts on the agitation term $\delta \boldsymbol{F}_a(t)$ and duration $\delta t_a$ to $\delta \boldsymbol{F}_P(t)$ and $\delta t_P$ respectively. Pressure fluctuation agitations will be modeled as a series of randomly directed discrete events, and for simplicity, the magnitudes $\delta F_P$ and $\delta t_P$ are assumed constant.

The physical forms of $\delta F_P$ and $\delta t_P$ to be derived must obey the fluctuation-dissipation (F-D) theorem that appears in different guises in a variety of other contexts.[26] Most orthodox derivations of the theorem are mathematically involved and obscure the underlying physics of this important relationship.[27] The following heuristic derivation, which is off by a factor of 2, emphasizes the role of velocity fluctuations in restoring the thermal speed erased by friction.

Consider a BP that is initially traveling with its equilibrium thermal velocity $v_{th} = \sqrt{d k_B T / M}$. Let $\delta v$ be a random fluctuation in velocity from a single $\delta \boldsymbol{F}_P$ of duration $\delta t_P$. Neglecting the dissipation of each of a series of such random velocity fluctuations, they add together to form a random walk in velocity space. And like any random walk, the square of the net velocity change in time $t$ is proportional to the number of impulses received in that time. In time $\tau$ during which the initial thermal speed is partially dissipated, the number of agitation impulses received is then $N_\tau = \tau / \delta t_P$. To obtain good statistical results, we must assume that the number of events is large, so we must assume that $\delta t_P << \tau$. (This condition will be seen to hold very well for a particle of radius $10^{-6}$ m in air or water.) But the average squared speed of the BP must, by the equipartition theorem, stay at the constant value $v_{th,d}^2 = d k_B T / M$, as suggested by Figure 3. The squared speed obtained in time $\tau$ from the $N_\tau$ impulses is then $\delta v^2 N_\tau = \delta v^2 \left( \tau / \delta t_P \right)$. But this addition must replace the original squared thermal speed, so $\delta v^2 \left( \tau / \delta t_P \right) = v_{th,d}^2 = d k_B T / M$ This is not exactly right, of course, for two reasons. First, the initial thermal velocity dissipates exponentially and so is not dissipated completely in time $\tau$. And second, as mentioned, the dissipations of sequential increments of velocity are not accounted for, which tends to exaggerate their effects. A correct approach for infinite time and dissipation of velocity increments, given in the Appendix, shows that the factor $\tau / \delta t_P$ should be half as great. So we should have:

$$\delta v^2 N_\tau = \delta v^2 \left( \tau / 2 \delta t_P \right) = v_{th,d}^2 = d k_B T / M \ . \qquad (18)$$

By the impulse-momentum theorem, $\delta v^2 = \delta F_P^2 \delta t_P^2 / M^2$, so substituting for $\delta v^2$, we get a version of the F-D theorem that shows clearly the relationship that must hold between agitation and dissipation:

$$\delta v^2 \left( \tau / 2 \delta t_P \right) = \delta F_P^2 \tau \, \delta t_P / 2 M^2 = d k_B T / M \ . \qquad (19)$$

Substituting for $\tau$ from Eq. (15) and simplifying, we get the commonly stated form of the F-D theorem:

$$\delta F_P^2 \delta t_P = 12 d \pi \eta a k_B T \ . \qquad (20)$$

The derivation of Eq. (20) used the thermal equilibrium speed $v_{th,d}^2 = d k_B T / M$. In general, if $\delta F_a$ and $\delta t_a$ obey the F-D theorem, then we can be assured that the thermal equilibrium speed is implied. This is shown explicitly in the Appendix for $\delta F_P$ and $\delta t_P$.

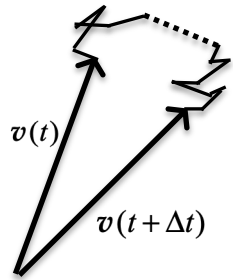

**Figure 3:** Many small random increments of velocity in velocity space add together over time to substantially change the direction of the velocity.

We will now derive expressions for pressure fluctuation and its associated relaxation time for d = 3, and show that they are consistent with Eq. (20) to within a small numerical factor. We follow the standard assumption that the volume of fluid exercising the effective pressure fluctuation force on the spherical BP is roughly the volume of the particle itself.[28] A pressure fluctuation in a parcel of fluid whose volume equals that of the Brownian particle can result in a number of ways: a fluctuation in the density at constant temperature,[29] which would result from a temporary influx or efflux of particles from a given volume; a thermal fluctuation at constant density;[30] or an adiabatic volume change in which density and temperature change. The simplest case is an adiabatic volume change. In fact, statistical mechanics shows that an adiabatic volume change is equivalent to a combination of the other two mechanisms.[31]

Consider a parcel of fluid of initial volume $V_i$.[32] For an adiabatic process assume pressure is changed from an initial value $P_i$ to a final value $P_f = P_i + \Delta P_f$, while the volume changes from $V_i$ to $V_f = V_i + \Delta V_f$. Let intermediate values during the change be $P = P_i + \Delta P$ and $V = V_i + \Delta V$. For $\Delta P_f$ and $\Delta V_f$ not too large, $\Delta P \approx -\left(\partial P / \partial V\right)_S \Delta V$. The work associated with $\Delta P_f$ and $\Delta V_f$ is then:

$$\Delta W = \int_0^{\Delta V_f} \Delta P d(\Delta V) \approx \int_0^{\Delta V_f} -\left(\frac{\partial P}{\partial V}\right)_S \Delta V d(\Delta V) \approx -\left(\frac{\partial P}{\partial V}\right)_S \frac{\Delta V_f^2}{2}. \quad (21)$$

From $\Delta P_f \approx -\left(\frac{\partial P}{\partial V}\right)_S \Delta V_f$, $\Delta V_f^2 = \Delta P_f^2 / \left(\partial P/\partial V\right)_S^2$, so Eq. (21) is:

$$\Delta W = -\frac{1}{2}\frac{\Delta P_f^2}{\left(\partial P/\partial V\right)_S}, \quad (22)$$

so:

$$\Delta P_f^2 = -2\left(\partial P \big/ \partial V\right)_S \Delta W. \quad (23)$$

Therefore:

$$\delta P_f \equiv \sqrt{\Delta P_f^2} = \sqrt{-2\left(\frac{\partial P}{\partial V}\right)_S \Delta W}. \quad (24)$$

Since $\Delta W$ in Eq. (21) is quadratic in the volume change, the equipartition theorem applies, so $\Delta W = k_B T / 2$. Therefore, substituting for $\langle \Delta W \rangle$ in Eq. (24),

$$\delta P_f = \sqrt{-k_B T\left(\frac{\partial P}{\partial V}\right)_S} = \sqrt{\frac{B_S k_B T}{V}}, \quad (25)$$

where $B_S = -V\left(\partial P / \partial V\right)_S$ is the adiabatic bulk modulus. The simplest guess for the net pressure fluctuation force on the Brownian particle is then $\delta P_f$ times the cross-sectional area of the particle, $\pi a^2$, where $\delta P_f$ is calculated from Eq. (25) for a fluid element of volume equal to the particle's volume.[33] Multiplying $\delta P_f$ by $\pi a^2$, substituting $4\pi a^3 / 3$ for $V$, and squaring, we get $\delta F_P^2 = 3\pi a B_S k_B T$, which we take as an approximation for $\delta F_P^2$ in Eq. (20). This is roughly correct, but it is easy to imagine an increase in pressure on one side of a particle and a decrease on the other, with parcels in other directions balancing, in which case the total force on the particle would be twice $\delta P \pi a^2$. In addition, pressure fluctuations occur at all scales, bigger and smaller than the volume of the particle. Like Kivelson et al., a correction constant $\kappa$ will be used to account for the fact that the estimate is not rigorous:[34]

$$\delta F_P^2 = \kappa^2 \left(3\pi a B_s k_B T\right). \qquad (26)$$

The next step is to determine the value of the relaxation time $t_P$ for a pressure fluctuation. First consider a volume of fluid and imagine that an external force is abruptly applied to it. This will change the pressure, but the change does not occur instantaneously. Instead, there is a very brief delay associated with the time required for the molecules in the volume to adjust appropriately. This relaxation time is essentially the time constant for an exponential decay of excess stress inside a material, and was first examined by J.C. Maxwell who corrected the false assumption that a fluid could not support a shear force.[35]

An expression for $\delta t_P$ appropriate for liquids can be determined by first considering a dense gas such as air. When a force is abruptly applied to a parcel, molecular adjustment is achieved by collisions, and the relaxation time in this case is just the mean time between collisions: $\delta t_P = L/v_{th}$, where $L =$ mean free path, $v_{th} =$ equilibrium thermal speed $= \sqrt{3k_B T/m}$, and $m =$ molecular mass. This estimate is fine for a gas like air at STP, because the mean free path is approximately 100 times the molecular diameter, making the concept of mean free path meaningful. This is not so in a liquid, where the distance between collisions of the crowded molecules is the same order of magnitude as, or even smaller than, the molecular diameter. But the simple expression $\delta t_P = L/v_{th}$ for a gas can be put into a form involving physical parameters that will work equally well for liquids and even solids, as follows.

Simple kinetic theory shows $L$ is connected to the viscosity of the gas:[36] $L = \dfrac{\eta}{nm\overline{v}}$, where $\overline{v} =$ (arithmetic) average speed $= \sqrt{8k_B T/\pi m}$ and $n =$ number density of the molecules. This leads to $\delta t_P = \dfrac{1.18\eta}{nk_B T}$. But the bulk modulus at constant entropy, $B_S$, is simply connected to the bulk modulus at constant temperature, $B_T$ : $B_S = \gamma B_T$, where $\gamma \approx 1$ is the adiabatic constant, i.e., the ratio of specific heat at constant pressure to that at constant volume (and equal to 5/3 for a monatomic gas); and $B_T = -V\left(\partial P/\partial V\right)_T$. Using the gas law, $P = Nk_B T/V$, and calculating $B_T$, we eliminate $nk_B T$ from $\delta t_P = \dfrac{1.18\eta}{nk_B T}$, which gives $\delta t_P = 1.08\gamma\eta/B_S \approx \eta/B_S$. Since viscosity and bulk modulus are easily measured properties of all fluids, we follow a number of authors and take $\delta t_P$ to be:[37]

$$\delta t_P = \frac{\eta}{B_s}. \qquad (27)$$

Now when this value of $\delta t_P$ is multiplied by $\delta F_P^2$ in Eq. (26), we find exact equivalence with Eq. (20) for d = 3 subject to the condition $\kappa^2 = 12$ so:

$$\kappa = 2\sqrt{3}. \qquad (28)$$

The validity of this value will be checked in the Appendix.

Table 1 give values of $\eta$, $B_S$, $t_P$, and the self-diffusion constant, $D_{mol}$, for molecules of water and air at STP. The $D_{mol}$ values are used to calculate the root-mean-square displacement $r_{rms}^{mol}$ of a molecule over the time $\delta t_P$ in three dimensions using Eq. (6) to test if the assumption of a fixed fluid parcel is valid. For water, $r_{rms}^{mol} = \sqrt{6D_{mol}t_P}$ is much smaller than the diameter of a BP of radius of $10^{-6}$ m. This validates the assumption implied in Eq. (25) for pressure fluctuations that the number of molecules in a fluid parcel of BP size doesn't change significantly over the time $\delta t_P$, i.e., that we are dealing with a fixed parcel of fluid. For air, $r_{rms}^{mol}$ is about 5% of the diameter of a particle of radius $10^{-6}$ m, so the assumption of a stable parcel is acceptable. The density of the BP is assumed to be twice that of water.

**TABLE 1: Physical Characteristics of Water and Air** ($a = 10^{-6}$ m, $M = 8.34\text{x}10^{-15}$kg )

|  | $\eta$ , kg/sm | $B_S$, Pa* | $\delta t_P$, s | $\tau$, s | $D_{BP}$, $m^2/s$ | $D_{mol}$, $m^2/s$ ** | $r_{rms}^{mol}(t_P)$, m |
|---|---|---|---|---|---|---|---|
| Water | 0.001 | $2.1\text{x}10^9$ | $4.76\text{x}10^{-13}$ | $2.12\text{x}10^{-7}$ | $2.20\text{x}10^{-13}$ | $2.02\text{x}10^{-9}$ | $5.34\text{x}10^{-13}$ |
| Air | $1.73\text{x}10^{-5}$ | $1.41\text{x}10^5$ | $1.24\text{x}10^{-10}$ | $1.23\text{x}10^{-5}$ | $1.27\text{x}10^{-11}$ | $1.85\text{x}10^{-5}$ | $1.17\text{x}10^{-7}$ |

* $B_s$ for water taken from A. Hudson & R. Nelson, *University Physics* (2nd ed., Sunders College Publishing, Philadelphia, 1990), p. 360; $B_s$ for air calculated from ideal gas law as $\gamma P$ , with $\gamma$ =1.4.

**Value for water taken from online article " https://dtrx.de/od/diff/index.html. Value for air taken as the experimental value for nitrogen from F. Reif, *Fundamentals of statistical and thermal physics* (McGraw-Hill, New York, 1965), p. 485.

From Table 1, $\delta t_P \ll \tau$, so the time-scales for agitation and dissipation are clearly well-separated. Let $\delta v_P$ be the random change in particle velocity from a single pressure-fluctuation event. From the impulse-momentum relationship $\delta F_P \delta t_P = M \delta v_P$ and neglecting the value of the correction constant $\kappa$, we get an approximate value of $\delta v_P$ for water of $5\times10^{-7}$ m/s, and for air about $10^{-5}$ m/s. After the time $\tau$ an initial equilibrium thermal velocity $v_{th}$ has significantly dissipated. In this time, the random impulses restore the thermal speed, so for water it takes about $\tau / \delta t_P = 225,000$ impulses, and for air about 50,000 impulses.

## V. FAILURE OF PRESSURE FLUCTUATIONS FOR LARGE VISCOSITY

The derivation of Eq. (20) depends on the assumption that $\delta t_P \ll \tau$ to insure the linear independence of dissipation and fluctuation terms in the Langevin equation. But as Table 2 shows, the assumption fails for glycerin for which $\eta$ is 1500 times water's, $\delta t_P$ is about 700 times water's, while $\tau = 1.48\text{x}10^{-10}$ s is much smaller than water's. The seriousness of the violation of the condition $\delta t_P \ll \tau$ follows when we consider a BP that starts with zero velocity. Then pressure fluctuations cannot bring the BP to its thermal speed of 0.00122 m/s, because even if a pressure fluctuation $\delta F_P$ acts for an infinite time (i.e., $t_P = \infty$ ) in contrast to a series of random fluctuations, the particle achieves a terminal speed $v_{term}$ (when $\delta F_P = 6\pi\eta a v_{term}$ ) of only $v_{term} = 9.48\text{x}10^{-4}$ m/s, about half its thermal speed. This same exercise shows that, for water, a constant pressure fluctuation would accelerate the BP to a terminal speed of 0.48 m/s, far above its expected thermal speed of . (Of course, a fluctuating random pressure gives it precisely its thermal speed.) We must conclude that pressure fluctuations are insufficient to overcome the drag effect of glycerin's very large viscosity.

A consistent treatment of the large-viscosity case requires the motion of the center of mass of a fluid parcel adjacent to the Brownian particle to be taken into account. Such a parcel has its own thermal velocity given by $v_{parcel} = \sqrt{3k_B T / M_{parcel}}$ .[38] Therefore, for a parcel of the same mass as the BP at rest, adjacent to and moving towards it, this coherent motion of molecules in the parcel exerts a large viscous force on the BP, thus forcing it to move in approximately the same direction as the parcel. In other words, there is an inelastic "collision" between the moving parcel and particle, which results in both having the same velocity after the interaction is complete. In the reference frame of the parcel, the particle plows into it and by Eq. (14), the particle is brought to rest relative to the parcel in a very short time $\tau$. This is also the effective lifetime of a parcel's impact on the BP. So for very large viscosity, when pressure fluctuations are largely irrelevant, Brownian motion is completely characterized by the single time scale $\tau$ for both drag and agitation force. Consequently, the motion is more like a random walk as in Figure 1(b) even at very small observation intervals.

**TABLE 2: Some Physical Characteristics of Glycerin**

| $\eta$ , kg/sm* | $B_S$ , Pa* | $\delta t_P$ , s | $\tau$ , s |
|---|---|---|---|
| 1.5 | 4.5x10$^9$ | 3.33x10$^{-10}$ | 1.48x10$^{-10}$ |

* $\eta$ taken from https://www.ecourses.ou.edu/cgi-bin/eBook.cgi?topic=fl&chap_sec=01.3&page=theory. $B_S$ taken as reciprocal of compressibility from http://hyperphysics.phy-astr.gsu.edu/hbase/Tables/compress.html.

It is interesting to look at how far into a parcel of glycerin the particle plows in a collision between the two. The coasting distance from Eq. (14) is:

$$\Delta_{coast} = \int_0^\infty v_o e^{-t/\tau}\, dt \approx \int_0^\infty v_{therm} e^{-t/\tau}\, dt \approx \tau v_{therm} = \frac{\sqrt{3Mk_B T}}{6\pi\eta a}. \qquad (29)$$

For a BP in glycerin, this is only 3.60x10$^{-13}$ m, about the size of an atomic nucleus. So the interaction between parcels and between parcels and BP is similar to a bunch of very rigid balls colliding with each other and with the BP; the fluid nature of the parcels is not so obvious.

A detailed treatment of an agitation force based directly on viscosity is beyond the scope of this paper, but the following reformulation of the Langevin equation presents a simple phenomenological model shedding insight into this aspect of Brownian motion.

Let the macroscopic motion of fluid enveloping the particle be described by an effective velocity $v_{parcel}$, measured relative to the lab frame as is $v$. Since the viscous force depends on the relative motion $v_{parcel} - v$ between the BP and the surrounding liquid, Newton's second law is:

$$M\frac{dv}{dt} = \delta F_P + 6\pi\eta a\left(v_{parcel} - v\right) = \left(\delta F_P + 6\pi\eta a v_{parcel}\right) - 6\pi\eta a v = \delta F_{total} - 6\pi\eta a v. \quad (30)$$

This is just the Langevin equation, showing the agitation force consisting of pressure fluctuations and fluid parcel motion. Since $v_{parcel} \approx v_{therm} = \sqrt{3k_B T/M}$, the criterion for pressure fluctuations to dominate is:

$$\frac{6\pi\eta a v_{parcel}}{\delta F_P} \approx \frac{6\pi\eta a v_{therm}}{2\kappa\sqrt{3\pi a k_B T B_S}} \approx \frac{\eta a}{\sqrt{a B_S M}} << 1, \qquad (31)$$

where factors of order unity have been ignored. Substituting for $M$ in terms of the fluid's mass density $\rho$, $M = 4\pi a^3 \rho / 3$, and using the fact that the speed of sound in a fluid is $v_s = \sqrt{B_S/\rho}$, we obtain the condition:

$$\eta << \rho a v_s \ . \qquad (32)$$

Applying this condition to a BP ($a = 10^{-6}$ m), the reader may easily verify that it holds well for water ($\rho = 1000$ kg/m$^3$, $v_s = 1400$ m/s, $\eta = 0.001$ kg/sm) but fails for glycerin ($\rho = 1261$ kg/m$^3$, $v_s = 1900$ m/s, $\eta = 1.5$ kg/sm).

## VI. SUMMARY

The diffusion rate of Brownian particles in relatively dense fluids like air at STP and water can be described by the Langevin equation with a Stokes form of friction involving collective viscous effects of the surrounding fluid. We have shown that the agitation term can be described by pressure fluctuations which also involve collective action of fluid surrounding the particle. The combination of Stokes friction and agitation from pressure fluctuations obeys the fluctuation-dissipation relationship. But the agitation model was shown to fail for very viscous fluids, taking glycerin as an example. In this case, the agitation results from both pressure fluctuations and from thermal motion of parcels of fluid surrounding the Brownian particle.

## APPENDIX

It was mentioned following Eq. (20) that fluctuation and dissipation processes obeying the F-D theorem will provide the BP with its thermal equilibrium speed as required by the equipartition theorem. To prove this, let $v_0$ be the initial velocity of the particle. As time progresses from $t = 0$ to a time $t = nt_P$, the particle is subject to $n$ randomly directed pressure fluctuations $\delta \boldsymbol{F}_{Pi}$ of duration $\delta t_{Pi}$, $i = 1$ to $n$, with $\delta \boldsymbol{F}_{Pi}$ and $\delta t_{Pi}$ taken to be of constant magnitude $\delta F_P$ and $\delta t_P$. For the $i^{\text{th}}$ impulse, the change in velocity is $\delta v_i = \delta \boldsymbol{F}_{Pi} \delta t_P / M$, and the magnitude of $\delta v$ from Eqs. (26) and (27) is:

$$\delta v = \left( \frac{\kappa \sqrt{3\pi a B_S k_B T}}{M} \right) \left( \frac{\eta}{B_S} \right) . \qquad (A1)$$

Then after the $n$ impulses, the velocity of the particle, accounting for the dissipation of velocity changes, is:

$$v(t) = v_0 e^{-t/\tau} + \sum_{i=1}^{n} \delta v_i e^{-(t-it_P)/\tau}, \quad t = nt_P . \qquad (A2)$$

Squaring and averaging over a large number of similar particles, and using the fact that the impulses are random in direction so $\left\langle \delta \boldsymbol{v}_i \cdot \delta \boldsymbol{v}_j \right\rangle = \delta v^2 \delta_{ij}$ ($\delta_{ij}$ is the Kronecker delta), we get:

$$\left\langle v^2 \right\rangle = \left\langle v_0^2 \right\rangle e^{-t/\tau} + \left\langle \delta v^2 \right\rangle e^{-2t/\tau} \sum_{i=1}^{n} e^{2it_P/\tau} . \qquad (A3)$$

The sum in Eq. (A3) is of the form $\sum_{i=1}^{n} \rho^i$, where $\rho = e^{2t_P/\tau}$. It is simply $\rho$ times the value of the geometric series $\sum_{i=1}^{n} \rho^{i-1}$ the sum of which is $\sum_{i=1}^{n} \rho^{i-1} = \frac{\rho^n - 1}{\rho - 1}$. So Eq. (A3) is:

$$\left\langle v^2 \right\rangle = \left\langle v_0^2 \right\rangle e^{-t/\tau} + \left\langle \delta v^2 \right\rangle e^{-2(t-t_P)/\tau} \left( \frac{e^{2t_P n/\tau} - 1}{e^{2t_P/\tau} - 1} \right). \qquad (A4)$$

Now assuming $t \equiv n t_P$ is $>> t_P$, i.e., that $n >> 1$, and remembering that $t_P << \tau$ for fluids like air and water, then $e^{2t_P/\tau} - 1 \approx 2t_P / \tau$. Then as $t \to \infty$, Eq. (A4) becomes:

$$\left\langle v^2 \right\rangle = \frac{\left\langle \delta v^2 \right\rangle \tau}{2 t_P}. \qquad (A5)$$

(The factor of 2 in the denominator explains the "fix" in the heuristic derivation leading to Eq. (20).) Substituting for $\tau$ from Eq. (15), $\delta v$ from (A1), and $t_P$ from Eq. (27), we get:

$$\left\langle v^2 \right\rangle = \frac{\kappa^2 k_B T}{4M}. \qquad (A6)$$

Enforcing the condition that this be equal to the equilibrium squared speed $\left\langle v^2 \right\rangle = \frac{3k_B T}{M}$, we get $\kappa = 2\sqrt{3}$, which is consistent with Eq. (28).

Note that $\left\langle v^2 \right\rangle$ in Eq. (A5) is proportional to $\left\langle \delta v^2 \right\rangle / t_P$. But $\delta v_i = \delta F_P \delta t_P / M$, so $\left\langle v^2 \right\rangle$ is proportional to $\delta F_P^2 t_P$, which is the left side of the F-D theorem in Eq. (20). Thus, when the F-D theorem is satisfied, the equilibrium speed will be $3k_B T / M$, as required by the equipartition theorem.

## REFERENCES

[1] Robert Brown, "A brief Account of Microscopical Observations made in the Months of June, July, and August, 1827, on the Particles contained in the Pollen of Plants; and on the general Existence of active Molecules in Organic and Inorganic Bodies," *Philosophical Magazine* N. S. 4 (1828), 161–173. Historical reviews can be found in Edward Nelson, *Dynamical Theories of Brownian Motion* (2nd ed.), Princeton University Press (1967), available online at https://web.math.princeton.edu/~nelson/books/bmotion.pdf; and Leonard B. Loeb, *The Kinetic Theory of Gases* (Dover, 1961, a republication of the 1934 2nd edition), Chapter VIII.

[2] The theory of Mössbauer spectra in gases, liquids, and solids was presented by K.S. Singwi and A. Sjolander, "Resonance Absorption of Nuclear Gamma Rays and the Dynamics of Atomic Motions," *Phys. Rev.*, **120**, 1093-1102 (1960) (reprinted in *The Mössbauer Effect*, Hans Frauenfelder ed. (W.A. Benjamin, New York, 1963)). One of the first experiments in liquids was done by Bunbury, D. St. P., *et al.*, "Study of diffusion in Glycerol by the Mössbauer effect of Fe57," *Physics Letters*, **6** (1), 15 Aug. 1963. This is one of many Mössbauer studies carried out in glycerin. Experiments by Keller, W. and W. Kündig, "Mössbauer studies of Brownian motion," *Solid State Comm.*, **16**, 253-256 (1975), used particles only 0.0165 microns.

[3] Robert, Aymeric, "Measurement of self-diffusion constant with two-dimensional X-ray photon correlation spectroscopy," *J. Appl. Cryst.* **40**, s34-s37 (2007). This technique is explained in great deal for Brownian particles by Clark, Noel A., *et al.*, "A Study of Brownian Motion Using Light Scattering," *Am. Jour. Phys.* **38** (5), 575-585 (1970). They use the technique to measure $D$ for particles of radius 0.063 microns.

[4] A. Einstein, *Ann. d. Phys.* **17**, p. 549 (1905). This is the first paper in A. Einstein, Investigations on the Theory of the Brownian Movement (edited with notes by R. Fürth, Dover Publications, New York, A.D. Cowper translator, 1956. Excellent summaries of Einstein's papers on the subject can be found in A.

Pais's great biography *Subtle is the Lord: The Science and the Life of Albert Einstein* (Oxford University Press, 1982), Chapter 5.

[5] These excellent historical reviews include this point: Piasecki, Jaroslav, "Centenary of Marian Smoluchowski's theory of Brownian motion," Acta Phys. Polonica, B, No. 5, **38**, 1623-1629 (2007); and Y. Pomeau and J. Piasecki, "The Langevin equation," *Comptes rendus-Physique*, **18** (9-10), pp. 570-582 (2017).

[6] Einstein's initial paper in Reference 1 was followed by a paper by Smoluchowski, M. (1906), "Zur kinetischen Theorie der Brownschen Molekularbewegung und der Suspensionen", *Annalen der Physik*, **21** (14): 756–780. This paper is based on the random walk. Einstein gave a random walk treatment in "The elementary theory of the Brownian motion," *Zeit. für Elektrochemie*, **14**, 1908, pp. 235-239. See also Mark Kac, "Random Walk and the Theory of Brownian Motion," *American Mathematical Monthly*, **54** (7), 369-391 (1947), reprinted in *Selected Papers on Noise and Stochastic Processes,* Nelson Wax (ed.), Dover Publications, New York (1954).

[7] Paul Langevin, ''Sur la théorie du mouvement brownien,'' C. R. Acad. Sci. (Paris) 146, 530–533 (1908), English translation introduced by Don S. Lemons, Anthony Gythiel translator, "Paul Langevin's 1908 paper on 'The Theory of Brownian Motion," *Am. J. Phys.* **65** (11), 1079-1081 (1997).

[8] The equipartition theorem is derived in many texts. See, e.g., Baierlein, Ralph, *Thermal Physics* (Cambridge University Press, New York, 1999), pp. 314-316; Schroeder, D.V., *An Introduction to Thermal Physics* (Addison Wesley Longman, New York, 2000), pp. 14-15, 28-29, and 238-240; and Sears, F., *Thermodynamics* (2ᵈ ed., Addison-Wesley, Cambridge, MA, 1955), p. 265pp. 297-298. A very simple derivation is given by de Grooth, B., *Am. J. Phys.* **67**, 1248-1252 (1999), but is not exactly correct: see Grassia, P., "Dissipation, fluctuations, and conservation laws," *Am. J. Phys.*, **69** (2), 113-119 (2001).

[9] On traps, see Mora, S. and Y. Pomeau, "Brownian diffusion in a dilute field of traps is Fickian but non-Gaussian," *Phys. Rev.* E, **98** (4), pp. 040101-1 to 040101-5 (2018); soft materials: Wang, B. *et al.*, "When Brownian diffusion is not Gaussian," *Nature Materials*, **11**, June 2012, pp. 481-485. For reviews of recent developments focusing on hydrodynamic effects, see Selmeczi, D., *et al.*, "Brownian motion after Einstein and Smoluchowski: Some new application and new experiments," *Acta Physica Polonica B*, **38** (8), pp. 2407-2431 (2007); Y. Pomeau and J. Piasecki, Ref. 5. These last two

[10] F.L. Markley and D. Park, "Microscopic and Macroscopic Views of the Brownian Motion," *Am. J. Phys.*, **40** (12), 1859-1860 (1972).

[11] D. Kivelson, S.J. Knak Jensen and M. Ahn, "Molecular theory of the translational Stokes-Einstein relation," *J. Chem Phys.* **58** (2), 425-433 (1973). (What we are calling Stokes friction is also known as Stokes-Einstein friction or relation.) Selmeczi *et al.* (Ref. 9) and Pomeau and Piasecki (Ref. 5) call for a collective model for agitation.

[12] Grassia, P., Ref. 8.

[13] Perrin's work is described in his comprehensive book *Atoms* (D. LL. Hammick, trans., Ox Bow Press, Woodbridge, CT, 1990 reprint of the original 1913 edition with 1921 Appendix.)

[14] E. Nelson, Ref. 1, pp. 68-70.

[15] See, e.g., M.F. Shlesinger *et al.*, "Above, below, and beyond Brownian motion," *Am. J. Phys.* **67** (12), 1253-1259 (1999). These authors say "a Brownian trajectory is infinitely jagged…Wiener proved that the distance between any two points on a Brownian trajectory is infinite, because the trajectory is actually two-dimensional, and not a simple curved line. We can avoid these mathematical difficulties of the Brownian trajectory if we consider the random walk motion to take place on a periodic lattice with jumps occurring at a regular rate."

[16] B. Lukić *et al.*, "Direct Observation of Nondiffusive Motion of a Brownian Particle," *Phys. Rev. Lett.*, **95**, 160601 (2005).

[17] Kheifets *et al*, *Science*, **343**, 1493-1496, 28 March 2014. This work is described in Mark G. Raizen and Tongcang Li, "The measurement Einstein deemed impossible," *Physics Today*, **68** (1), 56-57 (2015), and references therein. A good review is given by G. Volpe and G. Volpe, *Am. J. Phys.* **81**, 224-230 (2013).

[18] For example, Perrin compared displacements for data for observations every 30 s and every 120 s. As expected, the root mean squared displacements for the latter were $\sqrt{120/30} = 2$ times the former, i.e., the squared displacement was 4 times the former. See Loeb, Leonard B., *The Kinetic Theory of Gases* (Dover, 1961, reprint of 1934 edition), p. 403.

[19] Selmeczi *et al.*, Ref. 9.

[20] G.K. Batchelor, *An Introduction to Fluid Mechanics* (Cambridge Univ. Press, 1967), says (p. 149): "The validity of the no-slip boundary condition at a fluid-solid interface was debated for some years during the last [19th] century, there being some doubt about whether molecular interaction at such an interface leads to momentum transfer of the same nature as that as that at the surface in the interior of a fluid; but the absence of slip at a rigid wall is now amply confirmed by direct observation and by the correctness of its many consequences under normal conditions." A nice discussion of the no-slip condition is given by Berg, H.C., *Random Walks in Biology* (Princeton University Press, Expanded Edition, 1993), p. 54-55. Berg mentions the fact that the no-slip condition means the particle carries fluid with it.

[21] Small corrections of $f_{St}$ for air can be found in Perrin, *Atoms*, p. 185. These corrections were originally derived for the purpose of insuring that the charge of the electron, in Millikan's famous oil drop experiment, was as accurate as possible.

[22] Boussinesq, J. 1885, *C.R.Acad.Sci.,Paris* **100,** 935; Basset, A.B.,1888, *Phil.Trans.R.Soc.*A**179,** 43. The effect of the Boussinesq and Basset forces on the velocity autocorrelation function was determined by Vladimirsky, V. and Terletzky, Ya. 1945, *Zh. Eksp. Teor. Fiz.* **15,** 259. A good treatment using Laplace transforms is given by Lisy, V. and J. Tothova, "An old efficient approach to anomalous Brownian motion," an ArXiv reprint available at http://adsabs.harvard.edu/abs/2010arXiv1006.1060L.

[23] F. Reif, *Fundamentals of Statistical and Thermal Physics* (McGraw-Hill, New York, 1965), p. 572.

[24] See, e.g., G. Wannier, *Statistical Physics* (Dover, New York, 1966), p. 477. An alternative treatment is in D. Tabor, *Gases, Liquids, and solids* (Cambridge Univ. Press, 3rd ed., 1993), pp. 117-118.

[25] An interesting discussion of this condition is given in Manoliu, A., and C. Kittel, "Correlation in the Langevin theory of Brownian Motion," *Am. J. Phys.*, **47**(8), pp. 678-680 (1979).

[26] The theorem was formulated by H. Callen and T. Welton, "Irreversibility and Generalized Noise," *Phys. Rev.* **83**, 34 (1951). See also C.V. Heer, *Statistical Mechanics, Kinetic Theory, and Stochastic Processes* (Academic Press, 1972), pp. 440-442. An intuitive application of the theorem involves magnetic susceptibility. The response of a magnetic spin system to an external field is proportional to how rigidly the spins are coupled to each other, which then determines fluctuations in the average spin value. The less rigidity, i.e., the greater fluctuation, the greater is the response of the system to an external field, i.e., the greater the magnetic susceptibility. See, e.g., H. Eugene Stanley, *Introduction to Phase Transitions and Critical Phenomena* (Oxford Univ. Press, New York, 1971), Appendix A.

[27] Most standard derivations of the F-D theorem for Brownian motion are rather involved. For one-dimensional treatments, see F. Reif, *Fundamentals of Statistical and Thermal Physics* (McGraw-Hill, New York, 1965), pp. 570-572, and Daniel Gillespie, "Fluctuation and dissipation in Brownian motion," *Am. J. Phys.*, **61**(12), 1993. Grassia, P., Ref. 8, provides a proof based on elementary energy and momentum considerations. A rigorous but straightforward and concise treatment is given by P. Eastman, *Introduction to Statistical Mechanics*, Chapter 7, available online at https://web.stanford.edu/~peastman/statmech/friction.html. Eastman shows that the time scale for fluctuations (i.e., the correlation time for the fluctuating force) must be much smaller than the dissipation time. Another clear derivation is given by Dwight Neuenschwander, "Brownian Motion and the Fluctuation-Dissipation Theorem," SPS Observer (Winter 2004, pp. 10-14), originally printed in the Sigma Pi Sigma publication *Radiations* (Fall 2003, pp. 18-22).

[28] See, e.g., Markley and Park, Ref. 10, and Kivelson *et al.*, Ref. 11.

[29] Markley and Park, Ref. 10, suggest this particular course.

[30] See Kivelson *et al.,* Ref. 11, Appendix B, p. 433.

[31] A.M. Vasilyev, *An Introduction to Statistical Physics* (G. Leib translator, Mir Publishers, Moscow, 1983 revised from the 1980 Russian edition), pp. 326-327.

[32] What follows is adapted from P.M. Morse's treatment of density fluctuations in a gas. (P.M. Morse, *Thermal Physics* (Revised edition) (W.A. Benjamin, New York, 1964), pp. 228-229.)

[33] The arguments for using the volume of the particle are reasonable but in detail are subtle. See the discussion following Eq. (9) in Kivelson *et al.*, Ref. 11.

[34] See Kivelson *et.al.* (Ref. 11, Eq. (12)) They do not use the modulus $B_s$, but use their own modulus K which corresponds to fluctuations at constant density. (See last sentence of their Appendix B).

[35] J.C. Maxwell, "On the dynamical theory of gases," Phil. Trans. Roy. Soc. London, **157**, 57 (1867).

[36] Baierlein, Ralph, *Thermal Physics* (Cambridge University Press, New York, 1999), p. 364; Sears, F., *Thermodynamics* ($2^{nd}$ ed., Addison-Wesley, Cambridge, MA, 1955), p. 265, Eq. (13-12).

[37] Kivelson *et al.* (Ref. 11) say that the force correlation time can be associated with the time $4\eta/(3\rho c^2)$ where $c$ = speed of sound $= \sqrt{K_0/\rho}$, and $K_0 \equiv B_s$ is the "zero frequency bulk modulus." So their value is, apart from a small numerical factor, equal to $\eta/B_s$. A more precise result for $t_P$, but still within a small numerical factor of Eq. (27), can be found in P.A. Egelstaff, *An Introduction to the Liquid State* (Clarendon Press, Oxford, 1994), Chapter 4 and Appendix 5. Egelstaff gives two equivalent expressions for the force correlation time. In his notation, $\tau_m = \eta/G = \varsigma/B_1$, where the shear viscosity $\eta$ is paired with the "shear modulus," $G$, and the "bulk viscosity $\varsigma$ is paired with the bulk modulus $B_1$. But given the approximations in Kivelson *et al.*'s paper, $\varsigma \approx \eta$. Furthermore, bulk modulus values are more available than shear modulus values. For a good discussion of bulk modulus, or "second viscosity,' see Landau, L.D. and E.M. Lifshitz, *Fluid Mechanics* ($2^{nd}$ ed., Pergamon Press, New York, 1987), Section 81. See also: A.B. Bhatia, *Ultrasonic Absorption* (Dover Publications, New York, 1967)), pp. 50, ff and pp. 63-64; L. Hall, "The Origin of Ultrasonic Absorption in Water, *Physical Review, 75*, No. 7, 775-781; L.E. Kinsler and A.R. Frey, *Fundamentals of Acoustics* ($2^{nd}$ ed., John Wiley & Sons, New York, 1962), p. 224, Eq. (9.15a), and p. 235; R.W.B. Stephens and A.E. Bates, *Acoustics and Vibrational Physics* (Edward Arnold Ltd., London, $2^{nd}$ ed., 1966), Appendix 30 and Appendix 52. pp. 761, 763.

[38] E.M. Lifshitz and L.P. Pitaevskii, *Statistical Physics*, Part 2 (Pergamon Press, New York, 1980), p. 371 (Eq. 88.5); F. Munley, "Answer to Question #1 ["How does a Brownian particle at rest get kicked up to kT?," Frank Munley, *Am. J. Phys.* **62**(10), 871 (1994)], *Am. J. Phys.* **64**(3), 1996, p. 203.